\documentstyle[epsfig]{aipproc}

\newcommand{\muu}{\mu_{{\scriptscriptstyle\rightarrow} u}}
\newcommand{\mud}{\mu_{{\scriptscriptstyle\rightarrow} d}}
\newcommand{\thetad}{\theta_{{\scriptscriptstyle\rightarrow} d}}

\begin{document}
\title{Particle Acceleration at Ultra-Relativistic Shocks
       and the Spectra of Relativistic Fireballs}

\author{Yves A. Gallant$^{*,\dagger}$, Abraham Achterberg$^{*}$, \\
        John G. Kirk$^{\ddagger}$ and Axel W. Guthmann$^{\ddagger}$}

\address{$^{*}$Astronomical Institute, Utrecht University,
         Postbus 80\,000, 3508 TA Utrecht, Netherlands \\
         $^{\dagger}$Dublin Institute for Advanced Studies,
         5 Merrion Square, Dublin 2, Ireland \\
         $^{\ddagger}$Max-Planck-Institut f\"{u}r Kernphysik,
         Postfach 10\,39\,80, 69029 Heidelberg, Germany}

\maketitle

\begin{abstract}
We examine Fermi-type acceleration at relativistic shocks, and distinguish 
between the initial boost of the first shock crossing cycle, where the 
energy gain per particle can be very large, and the Fermi process proper 
with repeated shock crossings, in which the typical energy gain is of 
order unity. We calculate by means of numerical simulations the spectrum 
and angular distribution of particles accelerated by this Fermi process, 
in particular in the case where particle dynamics can be approximated as 
small-angle scattering. We show that synchrotron emission from electrons 
or positrons accelerated by this process can account remarkably well for 
the observed power-law spectra of GRB afterglows and Crab-like supernova 
remnants. In the context of a decelerating relativistic fireball, we 
calculate the maximum particle energy attainable by acceleration at the 
external blast wave, and discuss the minimum energy for this acceleration
process and its consequences for the observed spectrum.
\end{abstract}

\section*{Introduction}

   The spectrum of emission from GRB afterglows is well-accounted for by 
synchrotron emission from electrons accelerated at a decelerating 
relativistic blast wave.  The mechanism responsible for the 
gamma-ray-burst emission itself is less well-established, but
it has been interpreted as synchrotron emission as 
well, from electrons accelerated either at internal, mildly 
relativistic shocks, or at the ultra-relativistic external shock as it 
runs into a clumpy medium.

   In current models of afterglow emission, however (e.g.\
\cite{gallant:SarPirNar98}), the particle acceleration physics is
simply described by two parameters which are left to be 
adjusted to the observations: the shock is assumed to accelerate
the electrons to a power-law spectrum of index $p$, with a lower
cutoff $E_{\rm min}$ which is simply related to the efficiency of 
energy conversion into these accelerated electrons.  In what follows,
we first examine more closely the spectral index $p$ that can be 
theoretically expected for Fermi-type acceleration at relativistic 
shocks and compare it with observed values, and then consider
the maximum and minimum energies over which this spectrum can extend,
$E_{\rm max}$ and $E_{\rm min}$, and their consequences for 
observations.

\section*{Fermi acceleration at relativistic shocks}

\subsection*{Shock-crossing kinematics and energy gain}

   In what follows we restrict our attention to ultra-relativistic
shocks, i.e.\ those with Lorentz factor $\Gamma_{\rm sh} \gg 1$
with respect to the upstream medium.  For a weakly magnetised
shock, the shock jump conditions then imply a relative Lorentz factor 
$\Gamma_{\rm rel} \approx \Gamma_{\rm sh} / \sqrt{2}$ between the 
downstream and upstream media, and a shock velocity of $c/3$ relative 
to the downstream medium.

   Assuming a standard Fermi-type process at the shock, in which
charged particles are deflected elastically by magnetic fluctuations
in both the upstream and downstream media, the energy change of a
particle in a single cycle of crossing and re-crossing the shock
is given by
\begin{equation}
\frac{E_{f}}{E_{i}} = \Gamma_{\rm rel}^{2} ( 1 - \beta_{\rm rel}
\mud ) ( 1 + \beta_{\rm rel} \muu' ) .
\label{gallant:Egain}
\end{equation}
Here $E_{i}$ and $E_{f}$ are the particle's initial and final energies,
and $\mud$ and $\muu$ the cosine of its direction angle $\theta$ 
(between its velocity and the shock normal) upon crossing the shock
into the downstream and upstream media respectively.  Throughout, primed
and unprimed variables refer to quantities measured in the downstream
and upstream rest frames, respectively.

   Since kinematics require $1 \ge \muu' > \onethird$, the energy gain 
factor (\ref{gallant:Egain}) will depend most sensitively on the 
distribution of $\mud$.  For a pre-existing isotropic population of 
relativistic particles upstream, energy gains $E_{f}/E_{i}$ of order 
$\Gamma_{\rm rel}^{2}$ can be achieved in the first shock crossing 
cycle.  However, for particles having crossed the shock from
downstream, realistic deflection processes  upstream yield
$\thetad \lesssim 2 / \Gamma_{\rm sh}$, so that for all subsequent
shock crossing cycles, on average the particle energy is only
roughly doubled by each cycle \cite{gallant:GalAch99}.

\subsection*{Numerical simulations and the spectral index $p$}

   The power-law index of the accelerated particle spectrum depends
on the average energy gain per shock crossing and on the return 
probability, the chance that a particle crossing downstream
will eventually recross the shock upstream.  Both these factors
are strong functions of the angular distribution of particles
crossing the shock, which as suggested by the above considerations is
highly anisotropic.  Thus the quasi-isotropic approximations current
in non-relativistic shock acceleration do not apply, and we turn
to numerical simulations.

   For simplicity, we focus our attention here on the case where 
both the upstream and downstream particle dynamics are dominated by 
scattering; in other words, we assume that magnetic fluctuations, 
possibly amplified by turbulence downstream, dominate the effect of 
the regular magnetic field for transport in the shock normal direction.
Since the nature of the particle transport is by assumption independent
of particle energy, we decoupled the dynamical problem from the energy
gains, and first computed a numerical approximation to the function
$f_{d}(\muu' ; \mud')$, the distribution of downstream egress angles
$\muu'$ for a given ingress angle, by Monte-Carlo simulation of the
downstream scattering process for a grid of $\mud'$ values.  Along
with a similarly obtained representation of the upstream dynamics,
$f_{u}(\mud;\muu)$, and the energy gain formula (\ref{gallant:Egain}), 
these constitute the necessary ingredients for a Monte-Carlo 
calculation of the accelerated particle distribution.

\begin{figure}[b!]
\centerline{\epsfig{file=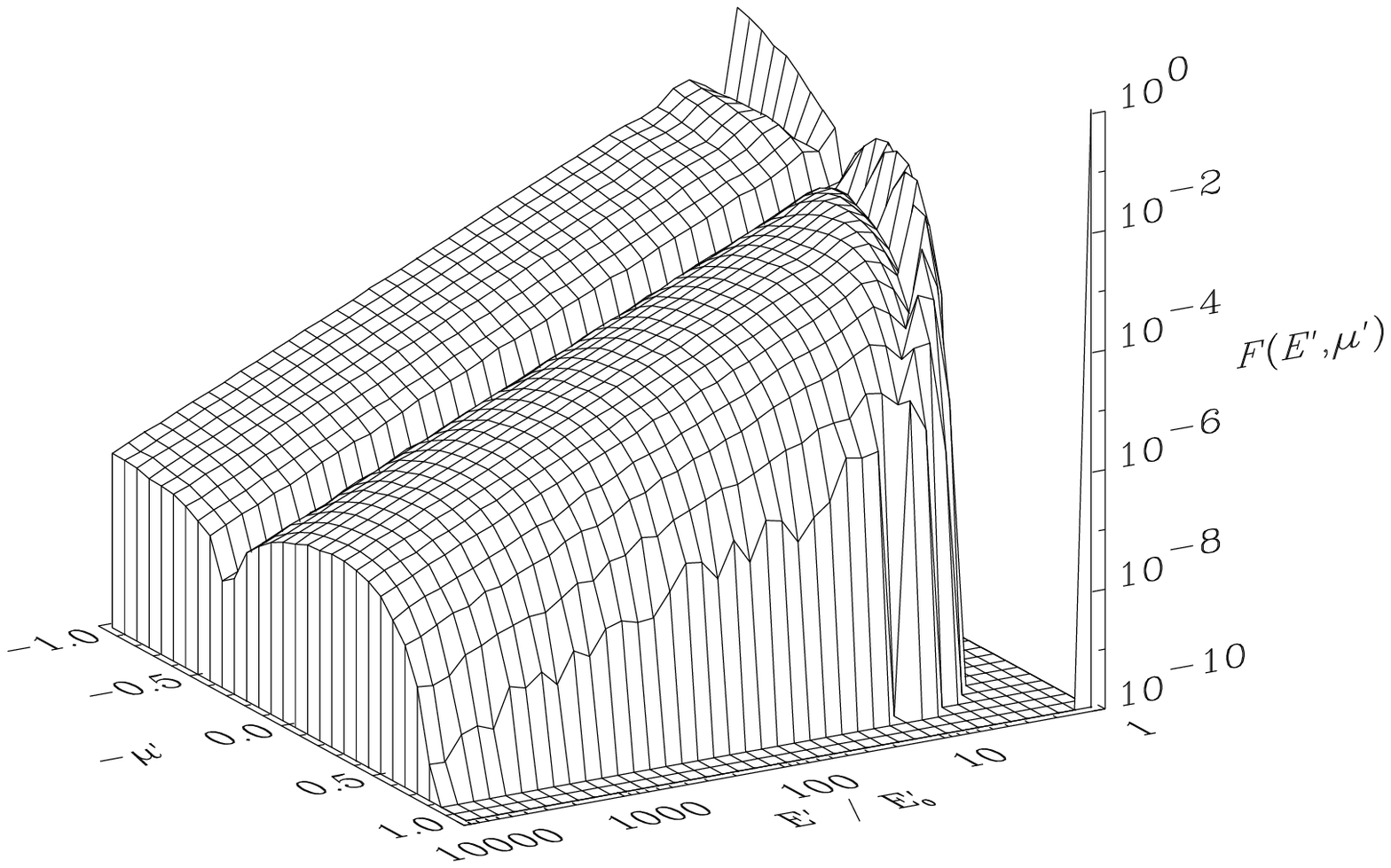,width=7.8cm}
            \epsfig{file=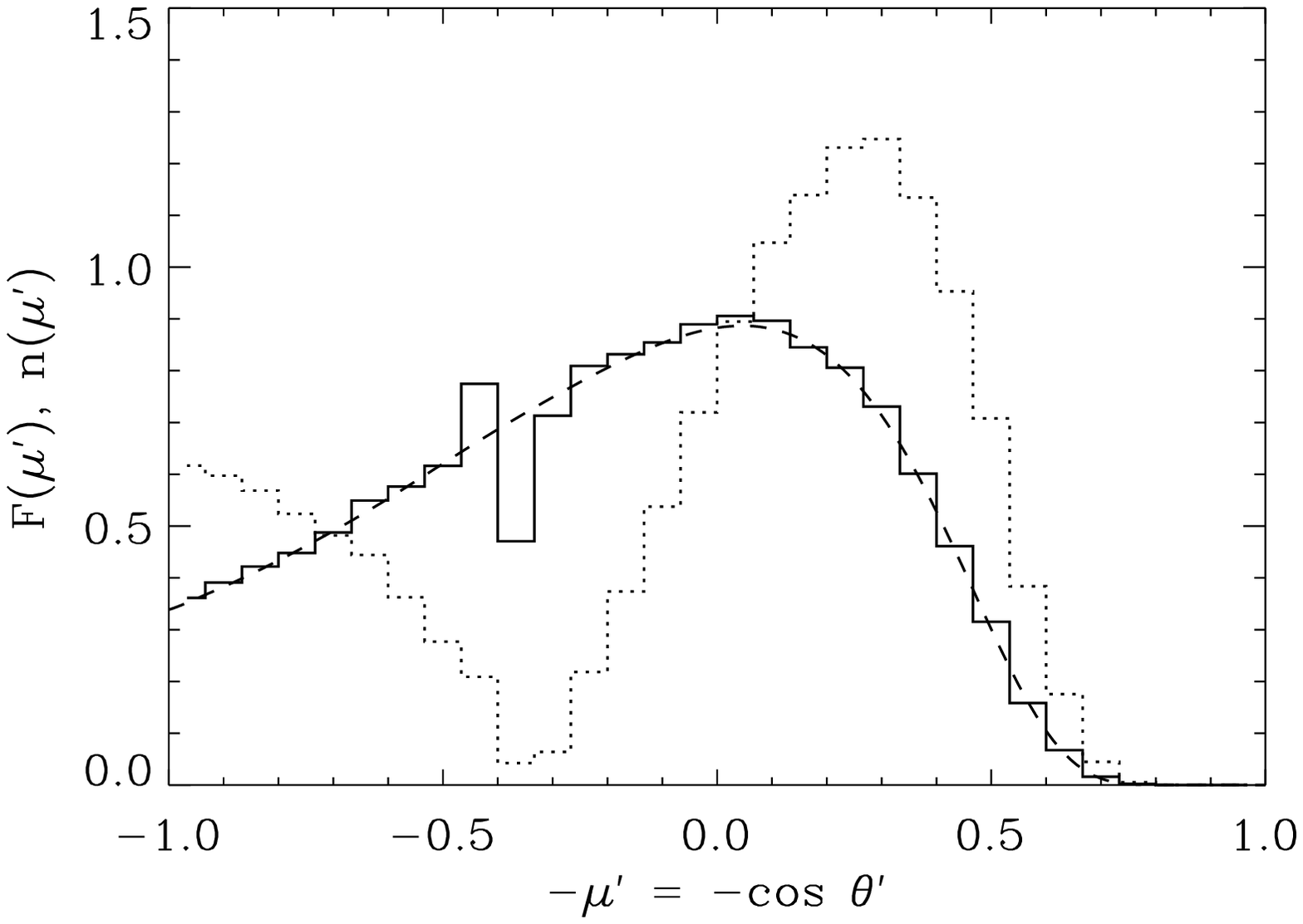, width=7.0cm}}
\caption{(a) Left panel: Steady-state flux distribution at the 
shock, $F(E', \mu')$, as a function of downstream particle energy 
$E'$ and direction angle cosine $\mu'$.  (b) Right panel: Asymptotic 
angular distribution of the particles at shock crossing, expressed both
in terms of flux $F(\mu')$ (dotted line) and density $n(\mu')$ (solid
line), each normalised to unity.  The dashed line shows the distribution
$n(\mu')$ obtained by the eigenfunction method for the same case.}
\label{gallant:Fig1}
\end{figure}

   The results of such a calculation are shown in Figure 
\ref{gallant:Fig1}(a).  Particles are injected at $E_{0}'$ with 
$\mud' = -1$, but the influence of this highly anisotropic initial
condition disappears after a little more than a decade in energy, where 
the self-consistent angular distribution is established with a smooth 
power-law dependence, $F(E', \mu') \propto F(\mu') E'^{-p}$.  We 
obtained $p = 2.23$ for this case, in perfect agreement with the
results of the semi-analytical eigenfunction method 
\cite{gallant:Gutetc00}.  The angular distribution obtained with 
that method is compared with the simulation results in Figure~%
\ref{gallant:Fig1}(b), which shows both the asymptotic flux 
distribution $F(\mu')$ measured in the simulations and the 
corresponding density distribution, $n(\mu') \propto F(\mu') / 
(\mu' - \onethird)$.  The agreement is excellent except near
the loss cone, $\mu' \approx \onethird$, where the denominator
amplifies Poisson noise in the measured Monte-Carlo flux.

   While the spectral index found above is valid only for  pure 
scattering, simulations incorporating more complex transport 
dynamics, especially upstream where the effect of the regular
magnetic field may well not be negligible, yield similar or only
slightly steeper spectra in the same ultra-relativistic limit
\cite{gallant:BedOst98,gallant:GalAchKir98}.

\subsection*{Comparison with observations}

   Early observations of two GRB afterglows suggested
$p = 2.3 \pm 0.1$ \cite{gallant:Wax97}, and detailed
analysis of the GRB 970805 afterglow spectrum yielded $p=2.2$
\cite{gallant:Galetc98}.  While multi-wavelength spectral
analyses have not been carried out in such detail for other 
afterglows, a value of $p \approx 2.2$ seems compatible with
most \cite{gallant:Fra99}.
Moreover, in another class of astrophysical objects where
an ultra-relativistic shock is thought to accelerate particles,
namely Crab-like supernova remnants, the inferred spectral indices
are similar: the best-fit model for the Crab Nebula spectrum 
corresponds to $p$ in the range $2.2$--$2.3$ \cite{gallant:KenCor84}.

This spectral index might thus be considered a signature of 
ultra-relativistic shock acceleration, as the spectral index for
acceleration at mildly relativistic shocks is expected to be
different \cite{gallant:Gutetc00}.
In this connection, it is intriguing to note that for GRB prompt 
emission, the average value of 2.12 for the high-energy spectral
index \cite{gallant:Preetc98}, when interpreted as cooled synchrotron
emission, corresponds to $p = 2.24$, in excellent agreement with the
theoretical
value found above.  This could perhaps be viewed as spectral evidence
that the prompt GRB emission also originates at an ultra-relativistic
shock, such as the external shock, rather than a mildly relativistic
one, such as internal shocks.

\section*{Maximum and minimum energies}

\subsection*{Age limit and ultra-high-energy cosmic rays}

   We first consider the maximum particle energy attainable by Fermi 
acceleration at a relativistic blast wave in the absence of energy
loss processes; this is set by the constraint that the acceleration
time $t_{\rm acc}$ be shorter than the age of the system.  Since the 
fractional energy gain per shock crossing cycle is typically of order 
unity, the acceleration time is roughly the cycle time, which is the 
sum of the upstream and downstream residence times, $t_{u}$ and 
$t_{d}$.  When the downstream magnetic field is simply the 
shock-compressed upstream field, one can show that $t_{d} \sim 
t_{u}$; if the downstream field is amplified from this value, e.g.\ 
by turbulence, $t_{d}$ is correspondingly shorter, so that the
total cycle time is of order $t_{u}$.  Comparing this with the age
of the fireball yields a maximum energy which is attained at the
beginning of the deceleration phase, and has value
\begin{equation}
E_{\rm max}^{\rm (age)} \simeq 5 \times 10^{15} B_{-6} \left(
   \frac{{\cal E}_{52} \Gamma_{3}}{n_{0}} \right)^{\onethird}
   {\rm eV},
\label{gallant:Eage}
\end{equation}
where $B_{-6}$ is the upstream magnetic field, ${\cal E}_{52}$ the
isotropic fireball energy, $\Gamma_{3}$ its initial Lorentz factor
and $n_{0}$ the upstream density, respectively in units of microgauss,
$10^{52}$ erg, $10^{3}$ and cm$^{-3}$.  This upper limit has 
important implications for the hypothesis that ultra-high-energy 
cosmic rays might be produced in GRBs \cite{gallant:GalAch99}.

\subsection*{Synchrotron loss limit and spectral upper cutoff}

   While the above upper limit is appropriate for cosmic-ray protons,
for the electrons responsible for the observed afterglow
emission synchrotron losses must also be taken into account.  This
yields the additional criterion that the energy lost to synchrotron
radiation in time $t_u$ upstream and $t_d$ downstream must
be less than the energy gained per shock crossing cycle.  If the
downstream field is simply the compressed upstream one, synchrotron
losses upstream and downstream will be comparable; otherwise, the
downstream losses will dominate, so we need only consider the
latter.  Using this criterion yields another upper limit
$E_{\rm max}^{\rm (syn)}$, which is more stringent than
(\ref{gallant:Eage}) when the downstream magnetic field $B'$
exceeds about $0.1\, \Gamma_{3}^{10/9} \xi^{2/3}$\,G, with weak
dependences on the other fireball parameters ${\cal E}_{52}$ and
$n_0$, where $\xi$ is the field amplification factor above simple
compression, i.e.\ $B' \equiv \xi \sqrt{8}\, \Gamma_{\rm sh} B$.

   The maximum synchrotron photon energy emitted by electrons
of energy $E_{\rm max}^{\rm (syn)}$ will be roughly 150\,MeV in
the proper (downstream) frame, independently of the value of the
magnetic field.  This $B$-independence is a generic result
for acceleration times scaling like the gyrotime, with longer
$t_{\rm acc}$ yielding correspondingly lower maximum photon
energies.  Boosting to the observer's frame yields
\begin{equation}
h \nu_{c, \rm max}^{\rm (syn)} \simeq 150\;\Gamma_3\;{\rm GeV};
\end{equation}
the same result also holds for synchrotron-limited acceleration
at internal shocks.
This suggests that establishing the presence of a cutoff to the
synchrotron spectrum in the range of {\em EGRET} to TeV gamma-ray
energies could place direct constraints on the fireball Lorentz
factor.

\subsection*{Minimum energy and electron pre-acceleration}

   The Fermi process requires that particles feel the shock as a
discontinuity, which requires that their Larmor radius be larger
than the shock thickness, which is in turn roughly the
downstream thermal ion Larmor radius.  This sets a minimum
electron energy for Fermi acceleration $E_{\rm min}' \simeq
\Gamma_{\rm sh} m_i c^2$, where $m_i$ is the ion mass,
corresponding to an observed synchrotron photon energy
\begin{equation}
h \nu_{c,\rm min} \simeq 160\;\xi\,B_{-6} \Gamma_3^4\;{\rm keV}.
\end{equation}
It is interesting to note that for our fiducial parameters this
falls in the BATSE break energy range, although it is unclear
how the strong dependence on $\Gamma_3$, in particular, could
keep $h \nu_{c,\rm min}$ constrained to a narrow range of values.

   The energy $E_{\rm min}'$ exceeds by a factor $m_i / m_e$
that resulting from randomisation of the bulk upstream
electron energy, and electrons must thus be pre-accelerated
by some other process before Fermi acceleration can operate.
One candidate for this pre-acceleration mechanism is the resonant
ion cyclotron wave acceleration process \cite{gallant:Hosetc92},
which efficiently accelerates electrons over precisely the required
energy range, and yields harder power-law spectra than those obtained
above, providing a possible explanation for the BATSE low-energy
spectral indices.

\section*{Summary}

   We examined the particle energy gain per shock crossing cycle for 
Fermi-type acceleration at ultra-relativistic shocks, and found that 
while the initial shock crossing cycle can yield a very large energy 
gain, in all subsequent crossing cycles the particle energy on average 
roughly doubles.  We used Monte-Carlo simulations to obtain the 
spectrum of accelerated particles, and found a power-law spectral
index $p = 2.23$ for the case of small-angle scattering, in agreement
with the results of the semi-analytical eigenfunction method.
This value is  compatible with those inferred from observations of
GRB afterglows and  Crab-like supernova remnants, and might thus be
considered a signature of ultra-relativistic shock acceleration.

   For protons, the maximum energy attainable by Fermi acceleration
at a relativistic blast wave is set by the age limit, and is of order 
$5 \times 10^{15}$\,eV for typical fireball parameters.  For 
electrons, synchrotron losses can become the dominant limiting factor 
for moderately amplified downstream magnetic fields.  In that case 
the maximum observed synchrotron photon energy is independent of the
magnetic field, and is proportional only to the fireball Lorentz
boost factor.
Finally, electrons must be pre-accelerated to a minimum energy 
comparable with the thermal ions' before they can undergo Fermi 
acceleration; a good candidate for the pre-acceleration mechanism 
is the resonant ion cyclotron wave acceleration process.

\section*{}

   This work was supported by the Netherlands Foundation for Research 
in Astronomy (ASTRON) under project 781--76--014, and by the European 
Commission under TMR programme contract ERBFMRX--CT98--0168.

\end{document}